\documentclass[
aps,
prd,
showpacs,
preprint,
tightenlines,
nofootinbib,
floatfix]
{revtex4}
\usepackage{graphicx,
}
\begin{document}
%
\title{                Unbinned test of time-dependent signals\\
                        in real-time neutrino oscillation experiments}
%
%
\author{        E.~Lisi}
\affiliation{   Dipartimento di Fisica
                and Sezione INFN di Bari\\
                Via Amendola 173, 70126 Bari, Italy\\}
\author{        A.~Palazzo}
\affiliation{   Dipartimento di Fisica
                and Sezione INFN di Bari\\
                Via Amendola 173, 70126 Bari, Italy\\}
\author{        A.M.~Rotunno}
\affiliation{   Dipartimento di Fisica
                and Sezione INFN di Bari\\
                Via Amendola 173, 70126 Bari, Italy\\}

\begin{abstract}
Real-time neutrino oscillation experiments such as
Super-Kamiokande (SK), the Sudbury Neutrino Observatory (SNO), the
Kamioka Liquid scintillator Anti-Neutrino Detector (KamLAND), and
Borexino, can detect time variations of the neutrino signal,
provided that the statistics is sufficiently high. We quantify
this statement by means of a simple unbinned test, whose
sensitivity depends on the variance of the signal in the time
domain, as well as on the total number of signal and background
events. The test allows a unified discussion of the statistical
uncertainties affecting current or future measurements of
eccentricity-induced variations and of day-night asymmetries (in
SK, SNO, and Borexino), as well as of reactor power variations (in
KamLAND).
\end{abstract}
\medskip
\pacs{
14.60.Pq, 13.15.+g, 26.65.+t, 95.55.Vj} \maketitle

\section{Introduction}

Among the experiments that have established the phenomenon of
neutrino flavor oscillations (see, e.g., \cite{Revi} for a recent
review), several ones are able to accurately tag each neutrino
event in the time domain. In particular,  the Super-Kamiokande
(SK) experiment \cite{SKde} and the Sudbury Neutrino Observatory
(SNO) \cite{SNOd} can detect solar neutrino events in real time
through the Cherenkov technique, while  the Kamioka Liquid
scintillator Anti-Neutrino Detector (KamLAND) \cite{KLde} can
detect reactor neutrino events in real time through the
scintillation light. The Borexino experiment (in construction)
\cite{BORE} will also allow real-time observations of solar
neutrino events through a liquid scintillator technique.

In all such experiments, the separation of the global neutrino
data sample into signal ($S$) and background ($B$) cannot be
performed on an event-by-event basis, but only on a statistical
basis, providing the total number of events in each class:
\begin{equation}\label{NSNB}
  N=N_S+N_B\ .
\end{equation}
Therefore, within a generic time sequence of events,
\begin{equation}\label{timeseries}
  \{t_i\}_{i=1,\dots,N_S+N_B}\ ,
\end{equation}
the $i$-th one can be either a signal event or background event,
with probability $N_S/N$ or $N_B/N$, respectively. Testing time
variations of the signal is then made difficult by unavoidable
statistical fluctuations of the background around its average
level. This problem is often tackled by binning the events in time
intervals $\Delta t$, which should be shorter than the signal
variation timescale, but also long enough to allow a reasonable
statistical subtraction of the background. However, as it was
emphasized in \cite{Four,Faid}, one can also directly use the
experimental information in Eqs.~(\ref{NSNB}) and
(\ref{timeseries}) without
binning.%
\footnote{The works \cite{Four,Faid} were focussed on the Fourier
analysis of time variations induced by vacuum oscillation
solutions to the solar neutrino problem, which are currently ruled
out.}

In this paper, building upon Refs.~\cite{Four,Faid}, we discuss
the statistical significance of possible time variations of
signals in real-time experiments, by means of a simple unbinned
test based on the time sequence of events $\{t_i\}$. In Sec.~II we
introduce the test. In Sec.~III we apply it to the discussion of
seasonal variations induced by the Earth's orbital eccentricity on
the solar neutrino signal in SK, SNO, and Borexino. In Sec.~III we
discuss the statistical significance of possible day-night
variations induced by Earth matter effects on the solar neutrino
signal in SK and SNO. In Sec.~IV we discuss the possibility of
tracking the time variation of the total reactor neutrino flux in
KamLAND, with and without the background from geoneutrinos or from
a speculative georeactor. Finally, we draw our conclusions in
Sec.~V. Technical details about the statistical power of the test
and about its comparison with the Kolmogorov-Smirnov unbinned test
are given in Appendix~A and B, respectively.

\section{AN Unbinned test of time variations}

Let us consider a real-time neutrino experiment characterized by
an event rate
\begin{equation}\label{RSB}
  R(t) = S(t) + B\ ,
\end{equation}
where $S(t)$ is a generic time-dependent signal and $B$ is the
background, assumed to be constant (up to statistical
fluctuations). If the function $S(t)$ is known (e.g., from
theoretical predictions), it can always be cast in the form:
\begin{equation}\label{S(t)}
  S(t)=S\left( 1+ \frac{\sigma_S}{S}\,f(t)\right)\ ,
\end{equation}
where $S$ and $\sigma^2_S$ are the average value and the variance
of $S(t)$ over the detection time interval $T$:
\begin{equation}\label{S}
  S=\frac{1}{T}\int_0^T dt\,S(t)\ ,
\end{equation}
\begin{equation}\label{sigmaS}
  \sigma^2_S=\frac{1}{T}\int_0^T dt\,[S(t)-S]^2\ ,
\end{equation}
so that $f(t)$ has zero mean value and unit variance in the same
interval $[0,\,T]$. It is natural to ask then how well can one
prove experimentally that the signal varies in time, i.e., that
$S(t)$ is different from its average $S$ or, equivalently, that
$f(t)\neq 0$.

A simple ``unbinned'' test of the hypothesis $f(t)\neq 0$ consists
in showing that the quantity
\begin{equation}\label{beta}
\beta = \sum_{i=1}^{N_S+N_B} f(t_i)\ ,
\end{equation}
defined in terms of the real-time event sequence in
Eq.~(\ref{timeseries}), is statistically different from zero. If
the variance $\sigma_\beta^2$ can be estimated, then the
difference of $\beta$ from zero can be expressed in terms of
standard deviations $n_\mathrm{sd}$,
\begin{equation}\label{nsd}
n_\mathrm{sd} = \frac{\beta-0}{\sigma_\beta}\ .
\end{equation}
We show in Appendix~A that, for $\sigma_S/S\ll 1$, it is
\begin{equation}\label{beta+-}
  \beta\pm \sigma_\beta \simeq N_S\,\frac{\sigma_S}{S}\pm
  \sqrt{N_S+N_B}\ ,
\end{equation}
which implies the following estimate for $n_\mathrm{sd}$:
\begin{equation}\label{nsdfull}
n_\mathrm{sd} \simeq \frac{\sigma_S}{S}\frac{N_S}{\sqrt{N_S+N_B}}
\ .
\end{equation}

Equation~(\ref{nsdfull}) provides a useful test of the sensitivity
to time variations. It express the statistical sensitivity (number
of standard deviations $n_\mathrm{sd}$) in terms of a theoretical
quantity (the ratio $\sigma_S/S$ for the expected signal) and of
two experimental quantities (the total number of signal and
background events, $N_S$ and $N_B$), independently of
specific detector details.%
\footnote{Detector details are relevant for the evaluation of
experimental systematic errors, including systematic variations of
the background and of the detection lifetime or efficiency (which
can be relevant in some cases not considered here, e.g.,
short-period time variations \cite{Lomb}). Since such evaluation
is beyond our expertise, we refer to statistical errors only in
this paper.}
The proportionality to $\sigma_S/S$ is clearly expected, but the
dependence on $N_S/\sqrt{N_S+N_B}$ is not entirely trivial. A
similar dependence arises in the determination of the Fourier
components of a periodic neutrino signal \cite{Four}. In addition,
we show in Appendix B that such dependence arises also in the
Kolmogorov-Smirnov unbinned test, and can thus be considered as a
rather general scaling rule for the statistical sensitivity to
time variations.

Despite its simplicity, Eq.~(\ref{nsdfull}) can be useful to
understand several issues related to the  experimental detection
of time-varying neutrino signals, without performing numerical
simulations or data fitting. Representative cases will be
discussed in the next Sections. The reader is referred to
Appendix~A for the derivation of Eq.~(\ref{beta+-}), upon which
Eq.~(\ref{nsdfull}) is based.

\section{Testing eccentricity effects}

Within the current large-mixing-angle (LMA) solution to the solar
neutrino problem \cite{Revi}, the neutrino oscillation phase at
the Earth is completely averaged out (see, e.g., \cite{Pedr}, and
the only effect of the Earth's orbit eccentricity
$(\varepsilon=0.0167)$ is a variation of the signal with the
inverse square distance,
\begin{equation}
\label{Secc} S(t) = S\left(1+2\varepsilon\cos \frac{2\pi
t}{T}\right)+O(\varepsilon^2)\ ,
\end{equation}
where $T=1$~year and $t=0$ at perihelion. In this case it is
$\sigma_S/S=\sqrt{2}\,\varepsilon$, so that Eq.~(\ref{nsdfull})
can be written as
\begin{equation}
\label{necc}
n_\mathrm{sd}=\sqrt{2}\,\varepsilon\,\sqrt{\frac{N_S}{1+N_B/N_S}}\
.
\end{equation}
The above equation will be used to test the statistical
sensitivity to eccentricity effects in  SK, SNO, and Borexino.

\subsection{Application to SK}

In its first phase of operation (SK-I), the Super-Kamiokande
experiment has collected
\begin{equation}
N_S\simeq 22400
\end{equation}
signal events induced by solar neutrinos \cite{Lomb,MSmy}. The
corresponding signal-to-background ratio can be estimated through
the well-known plot showing the distribution of {\em all\/} events
as a function of $\cos\theta_\mathrm{sun}$ \cite{Fuku}. In this
plot the background is flat in $\cos\theta_\mathrm{sun}$, while
the signal is peaked at $\cos\theta_\mathrm{sun}=1$ and extends
down to $\cos\theta_\mathrm{sun}\simeq 0.5$. By adopting the
reasonable cut $\cos\theta_\mathrm{sun}\gtrsim 0.5$, we
graphically estimate from \cite{Fuku} that
\begin{equation}
N_B/N_S\simeq 2.5\  .
\end{equation}
We thus expect From Eq.~(\ref{necc}) a sensitivity to eccentricity
effects equal to (in unit of standard deviation):
\begin{equation}
n_\mathrm{sd}\simeq 1.9\ .
\end{equation}

Our estimate $n_\mathrm{sd}\simeq 1.9$ might seem too low as
compared with the SK-I official analysis \cite{MSmy}, where
evidence for eccentricity effects is claimed at $\sim\! 3\sigma$
level. However, we note that the best-fit eccentricity measured by
SK is a factor $\sim\! 1.5$ greater than the true one \cite{MSmy}.
This accidental upward shift ($\varepsilon_\mathrm{SK}\simeq 1.5\,
\varepsilon$) leads to a corresponding sensitivity enhancement,
$n_\mathrm{sd}\simeq 1.9\times 1.5=2.8$, not far from the $\sim\!
3\sigma$ detection level in Ref.~\cite{MSmy}. Therefore, it may be
argued that the intrinsic sensitivity of the SK-I data sample to
eccentricity effects is at level of $\sim\! 2\sigma$, and that the
$\sim\! 3\sigma$ evidence in Ref.~\cite{MSmy} may partly be due to
an accidental upward fluctuation of the signal.

\subsection{Application to SNO}

The Sudbury Neutrino Observatory experiment has reported physics
results for two main phases: SNO I (pure D$_2$O, 306.4 live days)
\cite{SNOC,SNOD,SNOA} and SNO II (D$_2$O plus salt, 254.2 live
days) \cite{SNON}. The total number of solar neutrino events for
SNO I+II is
\begin{equation}
N_S\simeq 5662\ ,
\end{equation}
including all charged current (CC), neutral current (NC), and
elastic scattering (ES) events. The background is very efficiently
rejected ($N_B\simeq 208$ events in phases I+II), leading to
\begin{equation}
 N_B/N_S\simeq 3.7\times 10^{-2}\ .
\end{equation}
Our estimate for the SNO I+II sensitivity to eccentricity effects
is then [Eq.~(\ref{necc})]:
\begin{equation}
n_\mathrm{sd}\simeq  1.74\ .
\end{equation}
Therefore, we argue that the SNO experiment should already be able
to see first indications for eccentricity effects at $>90\%$ C.L.%
\footnote{For one degree of freedom, $90\%$ C.L.\ corresponds to
$1.64\sigma$.}
If the $N_B/N_S$ ratio is kept at the low current value, a future
increase of the statistics by a factor of two (three) should bring
the sensitivity to the interesting level of about $2.5\sigma$
($3\sigma$) in the future.

\subsection{Application to Borexino}

\begin{figure}[t]
\vspace*{+0.0cm}\hspace*{-0.8cm}
\includegraphics[scale=0.93]{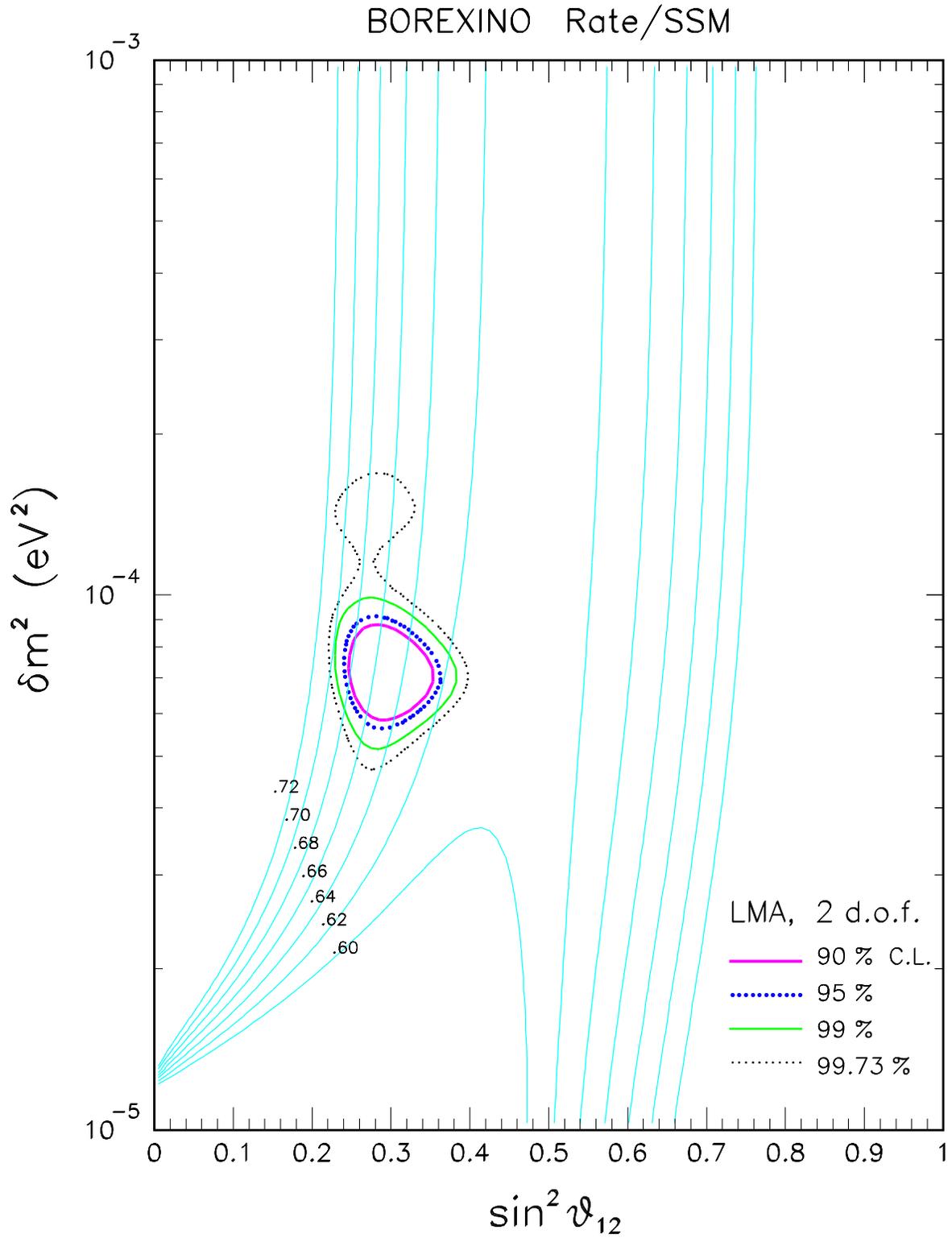}
\vspace*{-0.cm} \caption{\label{fig1}
\footnotesize\baselineskip=4mm  Suppression factor for the
Borexino event rate, superposed to the current LMA allowed region.
}
\end{figure}

The Borexino solar neutrino experiment (in construction)  is
expected to detect, in the absence of oscillations, about 55
events per day in the energy window $[0.25,0.80]$~MeV \cite{BORE}.
In the presence of oscillations, the rate suppression depends on
the values of the dominant oscillation parameters $(\delta
m^2,\theta_{12})$. Figure~1 shows isolines of the the suppression
factor in the mass-mixing plane, superposed to the current LMA
solution to the solar neutrino problem (as taken from
\cite{Ours}). From this figure we derive that the LMA-oscillated
event rate in Borexino should be roughly equal to 13000 events per
year, give or take one thousand events. After three years of data
taking, the number of signal events should then be around
\begin{equation}
 N_S\simeq 3.9\times 10^4\mathrm{\ (3\ years)}\ .
\end{equation}

One of the goals of the Borexino experiment is to reach a
background event rate comparable or smaller than the signal rate
\cite{BORE},
\begin{equation}
N_B/N_S\lesssim 1\ .
\end{equation}
If such goal is achieved, from Eq.~(\ref{necc}) we estimate that,
after three years of data taking, the sensitivity to eccentricity
effects should be definitely larger than $3\sigma$:
\begin{equation}
n_\mathrm{sd}\gtrsim 3.3\ .
\end{equation}
Such ``eccentricity test'' will convincingly prove that the
Borexino signal, despite the lack of directionality and the
nonnegligible background-to-signal ratio, does come from solar
neutrinos.

\section{Testing day-night asymmetries}

Within the current LMA solution to the solar neutrino problem (see
Fig.~1), Earth matter effects \cite{Matt} are expected to induce
marginal day-night variations of the signal $S$ in the SK and SNO
experiments (see, e.g., \cite{Pedr,MSmy,Petc}). Such effects,
which decrease with increasing $\delta m^2/E_\nu$ (where $E_\nu$
is the neutrino energy), should be null in Borexino.

Earth matter effects are usually parametrized in terms of a
day-night asymmetry $A$,
\begin{equation}\label{A}
A=\frac{S_n-S_d}{\frac{1}{2}(S_n+S_d)}
\end{equation}
where $S_n$ and $S_d$ are the average signal event rates during
night and day, respectively, with $S=(S_d+S_n)/2$. Within the LMA
solution it is $A>0$.

For our purposes, we can assume an approximate step-like variation
of the signal%
\footnote{Figure~2 in \cite{MSmy} shows, e.g., that the step-like
approximation in SK fails only for $0\lesssim \cos\theta_z
\lesssim 0.4$, which corresponds to a relatively small fraction
($\sim\! 1/5$) of the total signal.}
as a function of the zenith angle $\theta_z$ (a variable more
useful than the time $t$ in this context):
\begin{equation}\label{Sdn}
  S(\cos\theta_z)\simeq \left\{ \begin{array}{lll}
  S(1-A/2) &\; , \;  \cos\theta_z\in[-1,0] &
  \mathrm{(daytime)}\\
S(1+A/2) & \; , \; \cos\theta_z \in [0,+1] &
  \mathrm{(nighttime)}\\
  \end{array}\right.
\end{equation}
Within such approximation, the signal variance is simply given by
\begin{equation}\label{Sigmadn}
\frac{\sigma_S}{S}=\frac{A}{2}\ .
\end{equation}
Using Eq.~(\ref{nsdfull}), the statistical significance of a
day-night asymmetry $A$ is, in units of standard deviations,
\begin{equation}\label{nA}
n_\mathrm{sd}=\frac{A}{2}\sqrt{\frac{N_S}{1+N_B/N_S}}\ .
\end{equation}

We shall use the above equation in a slightly different version,
in order to estimate $\sigma_A$, namely, the $1\sigma$ statistical
uncertainty of $A$. This uncertainty is obtained by setting
$n_\mathrm{sd}=1$ and $A=\sigma_A$ in the above equation. The
result,
\begin{equation}\label{sigmaA}
  \sigma_A=2\,\sqrt{\frac{1+N_B/N_S}{N_S}}
\end{equation}
will be used for numerical estimates in SK, SNO, and Borexino.

\subsection{Application to SK}

As in Sec.~II~A, we take for SK-I the values $N_S\simeq 22,400$
and $N_B/N_S\simeq 2.5$. Our estimate for the $1\sigma$
uncertainty of the day-night asymmetry $A$ is then
\begin{equation}
\sigma_A\simeq 2.5\times 10^{-2}\ ,
\end{equation}
which is in reasonable agreement with the value $2.0\times
10^{-2}$ quoted by the SK Collaboration in Ref.~\cite{MSmy} for
the ``old'' analysis method, based on the comparison of integrated
daytime and nighttime samples.

In the same article \cite{MSmy}, the SK collaboration also
reported that, by using an extended maximum likelihood method, the
statistical uncertainty of the day-night asymmetry can be reduced
from $2.0\times 10^{-2}$ to $1.6\times 10^{-2}$, the latter being
adopted as official SK value for $\sigma_A$. We interpret this
reduction as an effective improvement in background rejection
achieved through the SK maximum likelihood method (i.e., lower
$N_B/N_S$ and thus lower $\sigma_A \propto \sqrt{1+N_B/N_S}$). We
cannot further elaborate on this interpretation for lack of
relevant information (the SK maximum likelihood analysis in
\cite{MSmy} is currently not reproducible outside the
Collaboration); public release of such information would be
beneficial to improve current global analyses of solar neutrino
data, and to allow combined SK+SNO tests of day-night asymmetry
effects (see also Sec.~IV~C).

\subsection{Application to SNO (phase I)}

\begin{table}
\caption{SNO data from phase I (306.4 live days): Total number of
events for each of the three classes CC, NC, ES and for the total
solar neutrino signal $S$ (2nd column), together with their
$\nu_e$ and $\nu_{\mu\tau}$ components for
$\phi_{\mu\tau}/\phi_e=1.94$ (3rd and 4th column, respectively).
See the text for details.}
\begin{ruledtabular}
\begin{tabular}{cccc}
Class & Total events& $\nu_e$ component & $\nu_{\mu\tau}$ component \\
$X$   &  $N_X$& $N_X^e$           & $N_X^{\mu\tau}$           \\
\hline
CC & 1967.7 & 1967.7 & 0 \\
NC & 576.5  & 196.3 & 380.2 \\
ES & 263.3  & 203.0 & 60.6 \\
\hline
$S$ & 2807.8 & 2367.0 & 440.8 \\
\end{tabular}
\end{ruledtabular}
\end{table}

In its first phase of operation \cite{SNOD,SNOA}, the SNO detector
has collected a total of $N_S=2807.8$ solar neutrino events with
small background ($N_B/N_S=4.4\times 10^{-2}$, neglected in the
following). The signal events are statistically separated into
three classes (CC, NC, and ES), each class containing different
contributions of events induced by $\nu_e$ and $\nu_{\mu,\tau}$.
Table~I shows the content of each class, as obtained by fixing the
neutrino flux ratio $\phi_{\mu,\tau}/\phi_e=3.41/1.76=1.94$ and
the cross-section ratio $\sigma_\mu/\sigma_e=1/6.48$ \cite{SNOD}.%
\footnote{Variations of the flux ratio
$\phi_{\mu,\tau}/\phi_e\simeq 1.94$ within experimental
uncertainties do not affect appreciably our results.}

The SNO-I result for the day-night asymmetry of the $\nu_e$
component of the total signal ($A_e$) is \cite{SNOA}
\begin{equation}\label{AeSNO}
A_e =
[7.0\pm4.9\mathrm{(stat.)^{+1.3}_{-1.2}\mathrm{(syst.)}}]\times
10^{-2}\ ,
\end{equation}
under the standard assumption of no asymmetry ($A=0$) for the
total neutrino flux $\phi=\phi_e+\phi_{\mu\tau}$. In this section
we reproduce with good accuracy the SNO-I statistical error
estimate $\sigma_{A_e}=4.9\times 10^{-2}$. On the basis of this
successful check, in the next Section we will try to estimate the
value of $\sigma_{A_e}$ after the SNO~I+II phases.

We remind that the $\nu_e$ and $\nu_{\mu\tau}$ components of the
solar neutrino signal in SNO can be separated only on a
statistical basis. Therefore, the $\nu_e$ asymmetry $A_e$ (which
is not directly observable) must be linked to the day-night
variation of the total signal $S$ (which is measurable), as
described in the following.

The constraint of no day-night change for the total flux
($\phi_e+\phi_{\mu\tau}$) implies that the (approximately
step-like) day-night variation of $\phi_e$,
\begin{equation}
\phi_e \to \phi_e\left(1\pm\frac{A_e}{2} \right)\ ,
\end{equation}
is associated to the following day-night variation of
$\phi_{\mu\tau}$,
\begin{equation}
\phi_{\mu\tau}\to\phi_{\mu\tau}\left(1\mp
\frac{A_e}{2}\,\frac{\phi_e}{\phi_{\mu\tau}}\right)\ .
\end{equation}
By applying the above variation factors to the $N_S^e$ and
$N_{S}^{\mu\tau}$ components of the total number of solar neutrino
events $N_S$, one gets the global day-night variation
\begin{equation}\label{NSAe}
N_S\to N_S\pm
\frac{A_e}{2}\left(N_e-N_{\mu\tau}\frac{\phi_e}{\phi_{\mu\tau}}\right)\
,
\end{equation}
which, using the numbers in Table~I (last row), provides the
desired link between $A_e$ and the total signal variation,
\begin{equation}
N_S\to N_S \left(1\pm 0.76\,\frac{A_e}{2}\right)\ .
\end{equation}
The analogous of Eq.~(\ref{sigmaA}) is then (for $N_S\simeq 2808$
and $N_B/N_S\ll 1$):
\begin{equation}
\sigma_{A_e} \simeq \frac{2}{0.76}\frac{1}{\sqrt{N_S}}= 5.0\times
10^{-2} \ ,
\end{equation}
in good agreement with the official SNO-I statistical error
\cite{SNOA}, also reported in Eq.~(\ref{AeSNO}). On the basis of
this successful test, we ``predict'' in the next section the
statistical error $\sigma_{A_e}$ expected from the day-night
analysis of SNO I+II data.

\subsection{Application to SNO (phase I+II)}

\begin{table}
\caption{SNO data from phases I+II ($306.4+254.2$ live days):
Total number of events for each of the three classes CC, NC, ES
and for the total solar neutrino signal $S$ (2nd column), together
with their $\nu_e$ and $\nu_{\mu\tau}$ components (3rd and 4th
column, respectively). See the text for details. }
\begin{ruledtabular}
\begin{tabular}{cccc}
Class & Total events& $\nu_e$ component & $\nu_{\mu\tau}$ component \\
$X$   &  $N_X$& $N_X^e$           & $N_X^{\mu\tau}$           \\
\hline
CC & 3307.3 & 3307.3 & 0 \\
NC & 1920.7  & 653.9 & 1266.8 \\
ES & 433.9  & 334.2 & 99.7 \\
\hline
$S$ & 5661.9 & 4295.4 & 1366.5 \\
\end{tabular}
\end{ruledtabular}
\end{table}

In its second phase of operation \cite{SNON}, the SNO detector has
increased the solar neutrino statistics by $2854.1$ events,
including 3307.3 CC, 1920.7 NC, and 433.9 ES events. Adding these
events to those of phase I, one gets the total numbers reported in
Table~II (where, for the sake of simplicity, we have used the same
$\phi_{\mu\tau}/\phi_e$ ratio as for the SNO phase I).

By inserting in Eq.~(\ref{NSAe}) the values of $N^e_S$ and
$N_S^{\mu\tau_S}$ from Table~II, we get the following day-night
variation for the total solar neutrino signal in SNO I+II
($N_S=5661.9$):
\begin{equation}
N_S\to N_S \left(1\pm 0.63\,\frac{A_e}{2}\right)\ ,
\end{equation}
so that
\begin{equation}\label{SNOerr}
\sigma_{A_e} \simeq \frac{2}{0.63}\frac{1}{\sqrt{N_S}}= 4.3\times
10^{-2} \ .
\end{equation}

Therefore, we estimate that the statistical error of the $\nu_e$
day-night asymmetry $A_e$ for SNO I+II should be about $4.3\times
10^{-2}$, i.e., not much smaller than for phase I only [see
Eq.~({\ref{AeSNO}})]. Notice that, if the CC sample were
hypothetically isolated on an event-by-event basis with no
background (an irrealistic goal), the corresponding statistical
uncertainty of the $\nu_e$ asymmetry would be
$\sigma_{A_\mathrm{CC}}=2/\sqrt{N_\mathrm{CC}}=3.5\times 10^{-2}$
for SNO I+II. These considerations suggest that, in any case, the
uncertainty of the $\nu_e$ day-night asymmetry in SNO I+II cannot
be smaller than $\sim\!4\times 10^{-2}$. This statement will be
checked soon, since the SNO I+II official day-night is currently
being finalized \cite{Grah}.%
\footnote{In the final analysis, the current SNO lifetime for
phase II  (254.2 days \cite{SNON}) might include additional 150
live days \cite{Grah}, i.e., about 60\% more statistics. In this
case, we estimate that the SNO~I+II error $\sigma_A$ should
decrease from $4.3\times 10^{-2}$ [Eq.~(\ref{SNOerr})] to
$4.0\times 10^{-2}$.}

\begin{figure}[t]
\vspace*{+0.0cm}\hspace*{-0.8cm}
\includegraphics[scale=0.86]{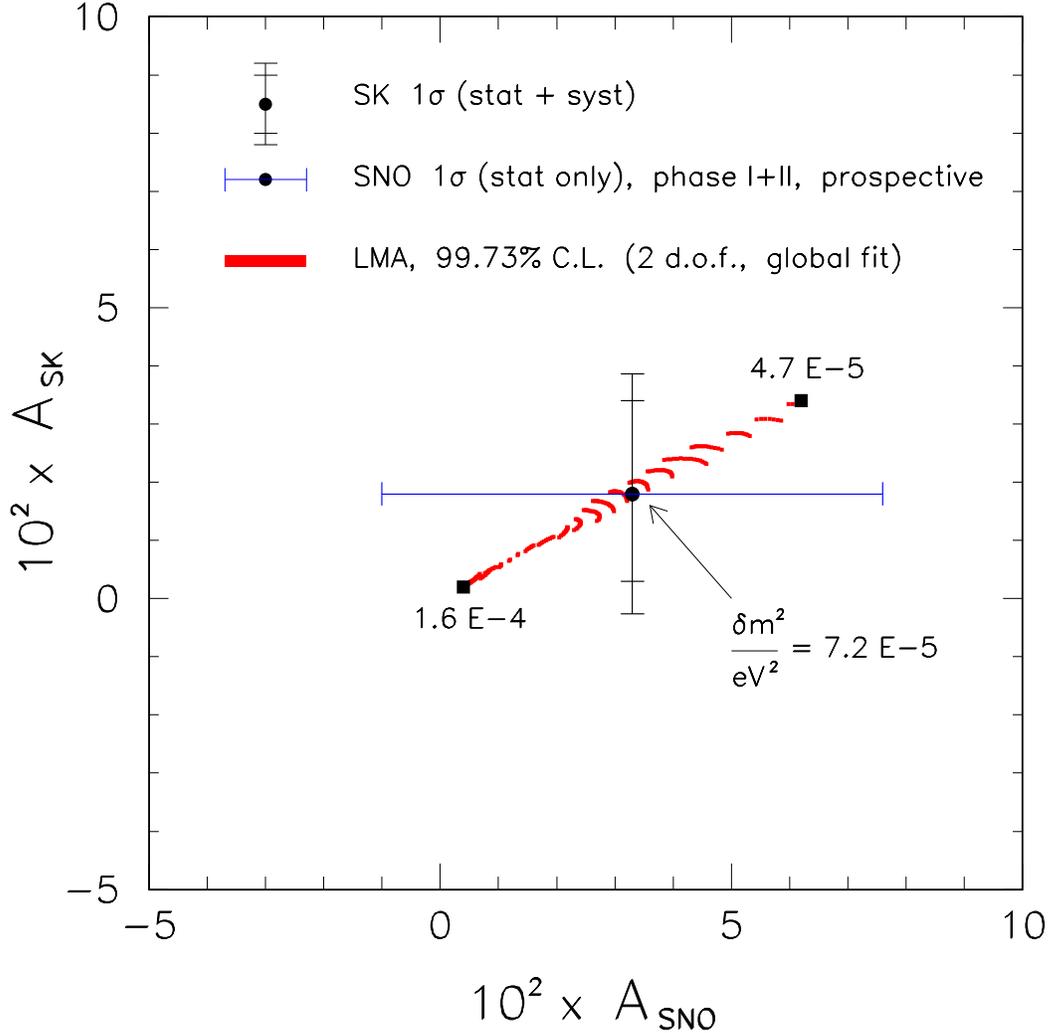}
\vspace*{-0.cm} \caption{\label{fig2}
\footnotesize\baselineskip=4mm  Correlation plot of the day-night
asymmetry in SK and in SNO. The slanted region maps the LMA
solution in Fig.~1 (at 99.73\% C.L.), which strongly correlates
the predictions for the day-night asymmetry in SK and SNO. Such
predictions depend mainly on $\delta m^2$ (three representative
values being shown in the figure). The datum with vertical errors
bars ($\pm1\sigma$ stat.+syst.) corresponds to the SK day-night
measurement from \cite{MSmy}. The ``datum'' with horizontal error
bars implements our estimate for the day-night asymmetry {\em
uncertainty\/} ($\pm1\sigma$ stat.\ only) from prospective
SNO~I+II data ($306.4+254.2$ live days), with hypothetical
best-fit central value.}
\end{figure}

Can the combination of the day-night asymmetries in SK (as
reported in Ref.~\cite{MSmy}) and in SNO I+II [our estimated error
in Eq.~(\ref{SNOerr})] improve the  current determination of the
LMA oscillation parameters? The answer can be derived from Fig.~2,
where we map the current LMA region at 99.73\% C.L.\ (as taken
from \cite{Ours}) onto the plane charted by the SK and SNO
day-night
asymmetries.%
\footnote{The ``substructures'' of the LMA region in Fig.~2
represent graphical artifacts, due to mapping of a finite number
of points.}
There is a strong positive correlations between the SK and SNO
asymmetries, as it was emphasized in Ref.~\cite{Corr} and
subsequently in Ref.~\cite{Kras}. The asymmetry rapidly decreases
for increasing values of the neutrino squared mass difference
$\delta m^2$, for which three representative values are shown in
Fig.~2 (including the LMA best fit, $\delta m^2=7.2\times
10^{-5}$~eV$^2$ \cite{Ours}). On top of the LMA region, we
superpose the SK day-night measurement at $\pm 1\sigma$ from
\cite{MSmy}, plus a prospective SNO I+II measurement characterized
by our estimate for the statistical error [Eq.~(\ref{SNOerr}) and
 no systematic error], and by the ``luckiest'' central value (on
top of the LMA best fit point). It can be seen that the SK+SNO
``error box'' at $\pm1\sigma$ is {\em larger\/} than the current
LMA region at 99.73\% C.L. Therefore, although the combination of
latest day-night asymmetry datum from SK \cite{MSmy} and of the
prospective one from SNO I+II (our estimate) can provide a useful
consistency check of the LMA predictions, we do not expect that
such data can significantly reduce the current LMA parameter
region.

\subsection{Application to Borexino}

In Borexino, one expects no day-night asymmetry ($A=0$) within the
LMA solution to the solar neutrino problem. Using
Eq.~(\ref{sigmaA}) and the same input values as in Sec.~II~C
($N_S=3.9\times 10^4$ and $N_S/N_B\lesssim 1$), we estimate  the
accuracy of this null result to be at least
\begin{equation}
A\pm\sigma_A = (0.0\pm 1.4)\times 10^{-2}\ ,
\end{equation}
after three years of data taking. A check of the null result
($A=0$) with percent accuracy  will provide a useful test that the
Borexino detector works as expected, although it will not improve
the determination of the neutrino oscillation parameters in the
LMA region.

\section{Testing reactor power variations in KamLAND}

The KamLAND experiment \cite{KLde,Land} is collecting events
induced by $\bar\nu_e$ produced in (mainly) Japanese reactors. The
reactor power demand (and thus the neutrino flux) in a given
country generally follows a seasonal trend \cite{Grat}. On top of
this trend, a strong temporary reduction of Japanese reactors'
power has occurred at the end of 2002 and in 2003 \cite{Hort}.
Figure~3 (graphically reduced from \cite{Hort}) shows the total
reactor fission flux during 540 days of KamLAND operation
(starting on March 4, 2002), where the first 216 days correspond
to the first phase with published results \cite{Land}. In the
following, we investigate whether such flux variations are
detectable with 540-day statistics,%
\footnote{We stop at 540 days since, to our knowledge, the reactor
flux at later times has not been publicly presented by the KamLAND
Collaboration so far.}
with or without background, by using Eq.~(\ref{nsdfull}).

\begin{figure}[t]
\vspace*{+0.0cm}\hspace*{-0.2cm}
\includegraphics[scale=0.88]{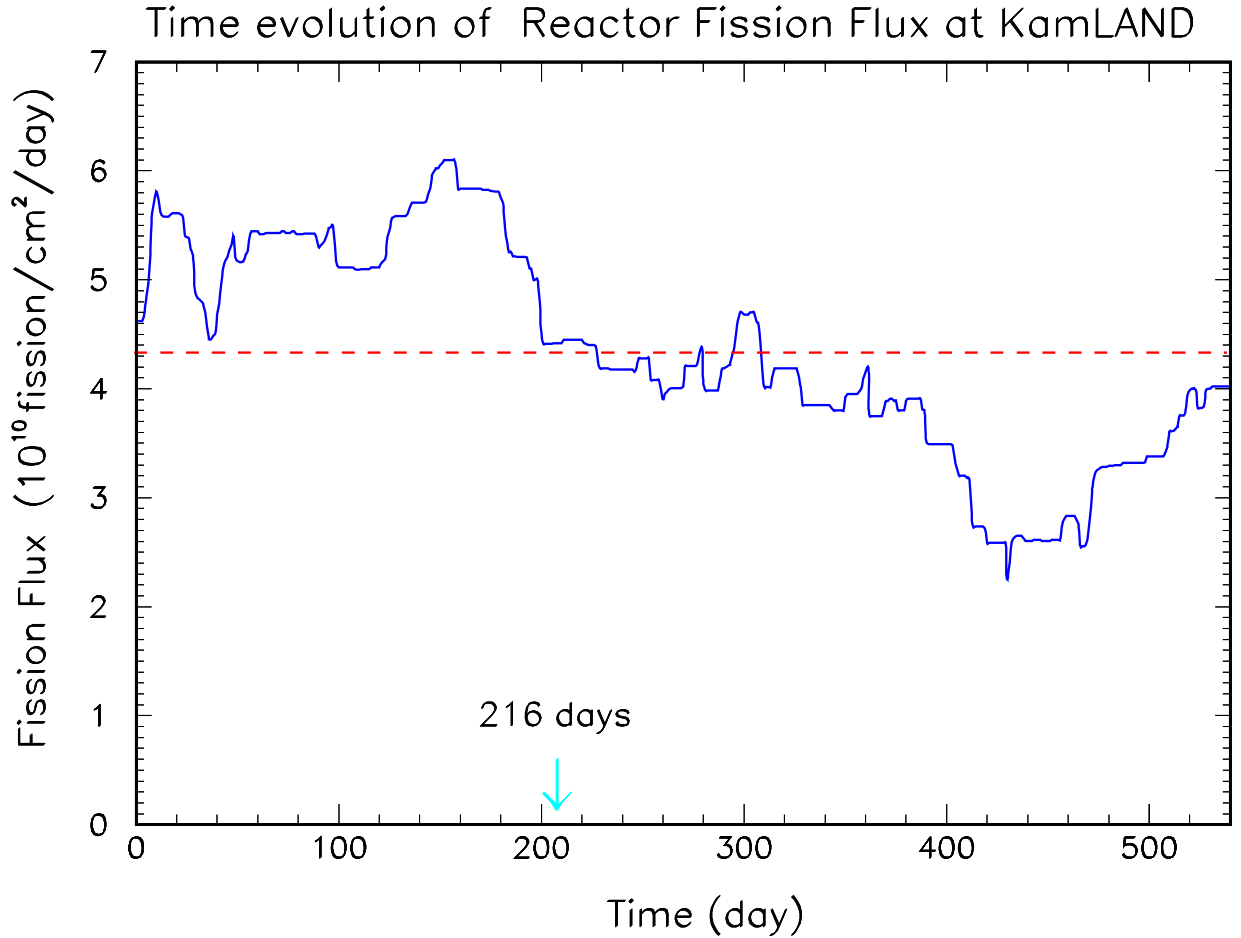}
\vspace*{-0.cm} \caption{\label{fig3}
\footnotesize\baselineskip=4mm  Variations of the total reactor
fission flux power in KamLAND during 540 days of operation (as
taken from \cite{Hort}. The horizontal dashed line represents the
average flux in the whole period. Published KamLAND data
\cite{Land} refer to the first 216 days.}
\end{figure}

\subsection{Case without background}

In the first 216 days, the KamLAND experiment has collected 54
events above the 2.6 MeV analysis threshold, which cuts away
basically all backgrounds \cite{Land}. If we naively rescale these
events using the time variation curve in Fig.~3, we estimate that
the statistics should be approximately doubled after 540 days,%
\footnote{Of course, a more detailed estimated should take into
account power variations and oscillation effects for each
reactor.}
\begin{equation}
N_S\simeq 2\times 54 = 108\ .
\end{equation}
From the same curve in Fig.~3, we obtain that
\begin{equation}
\label{sigmaK} \frac{\sigma_S}{S}\simeq 0.22
\end{equation}
over the whole period. Therefore, we estimate that the KamLAND
experiment should see an indication for reactor flux variations
with a statistical significance equal to
\begin{equation}
\label{nK}
n_\mathrm{sd}=\frac{\sigma_S}{S}\sqrt{N_S}\simeq 2.3
\end{equation}
standard deviations, after 540 days of operation.

\subsection{Case with background from geoneutrinos or from a georeactor}

Below the 2.6 analysis threshold, the main irreducible known
background in KamLAND is due to geoneutrinos \cite{Land}. Since
geo-$\bar\nu_e$ background predictions are uncertain  \cite{Fior},
we adopt for reference the KamLAND best-fit estimate of $N_B\simeq
9$ geoneutrino events, to be compared with the total reactor
signal with no threshold, $N_S\simeq 71$ events \cite{Land}.

The previous numbers refer to the first 216 days. At the end of
540 days of operation, the total reactor signal should be about
doubled (as noted previously),
\begin{equation}
N_S \simeq 142\ ,
\end{equation}
while the constant geoneutrino background should be rescaled by a
factor $540/216= 2.5$,
\begin{equation}
N_B \simeq 22.5\ .
\end{equation}
Inserting these numbers [and Eq.~(\ref{sigmaK})] in
Eq.~(\ref{nsdfull}), one gets
\begin{equation}\label{nK2}
n_\mathrm{sd}\simeq 2.4\ ,
\end{equation}
which is comparable to the no-background estimate in
Eq.~(\ref{nK}). Reactor flux variations can thus be tested at
$>2\sigma$ (after 540 days) also in the presence of
geo-$\bar\nu_e$ background.

While geoneutrinos represent a ``guaranteed''  $\bar\nu_e$
background, a natural reactor $\bar\nu_e$ source in the Earth's
core (``georeactor'') \cite{Hern} represents a more speculative
hypothesis, which is not endorsed by standard geochemical Earth
models (see, e.g., \cite{Dono}). For the sake of curiosity, we
estimate that --- as a rule of thumb --- a georeactor of power $p$
(in units of TW) should add a fraction of about $p\%$ to the
average KamLAND signal from man-made reactors. Assuming then a
maximum power $p\simeq 10$ TW \cite{Hern}, the additional
georeactor background ($N_B/N_S\simeq 0.1$) should degrade the
statistical significance estimated in Eqs.~(\ref{nK}) and
(\ref{nK2}) by $\sim\!0.1\sigma$ or less. We conclude that, with
or without geoneutrino or georeactor backgrounds, the KamLAND
experiment should definitely be able to see time variations of the
reactor signal at a level $\gtrsim 2\sigma$ (after 540 days of
operation). This check will provide further confidence on the
origin of the signal from reactor power plants.

If time variations of the reactor neutrino signal are successfully
tested in KamLAND, one could, in principle, turn the test around,
and try to constrain the amplitude of {\em constant\/} backgrounds
such as geoneutrino or georeactor events. This kind of tests,
together with detailed analyses of energy spectra, might improve
the discrimination of man-made reactor signals from geo-signals.
Such analyses could be performed also outside the Collaboration,
provided that: (1) the neutrino flux ``history'' of {\em each\/}
single reactor is publicly released, as it was emphasized in
\cite{Lisi}; (2) the energy and time tag of each event (background
and signal) is also released. The second condition is particularly
important to perform unbinned data analyses in the geoneutrino
energy window, where the few events being collected contain very
precious information for Earth sciences. We hope that the KamLAND
Collaboration will take these two desiderata into account.

\section{Summary and Conclusions}

Real-time neutrino oscillation experiments such as SK, SNO,
Borexino, and KamLAND, can detect time variations of the incoming
neutrino flux. We have introduced and discussed a possible test
for time variations, which does not require time-binning of the
events. The test provides a simple estimate for the statistical
sensitivity to time variations, in terms of the signal variance
and of total number of signal and background events
[Eq.~(\ref{nsdfull})]. This result has been used to discuss the
significance of eccentricity effects and day-night variations in
the SK, SNO, and Borexino solar neutrino experiments, as well as
the sensitivity to reactor power variations in KamLAND. In
particular, we estimate that:  (1) the combination of SNO I+II
data ($306.4 +254.2$ live days) can provide indications for
eccentricity effects at $>90\%$ C.L., but cannot provide a
determination of the day-night asymmetry accurate enough to
constrain significantly the current LMA allowed region of
parameters, even in combination with current SK day-night data;
(2) the Borexino experiment should test eccentricity effects at
$>3\sigma$ in three years; and (3) the KamLAND experiment should
already be able to test time variations of the reactor power at a
level $>2\sigma$. Such estimates are statistical only, and should
be somewhat degraded by systematic uncertainties, whose evaluation
is beyond the scope of this work.

\acknowledgments We thank Aldo Ianni for useful correspondence
about the Borexino experiment, and Daniele Montanino for helpful
comments. This work is supported in part by the Italian INFN
(Istituto Nazionale di Fisica Nucleare) and MIUR (Ministero
dell'Istruzione, Universit\`a e Ricerca) through the
``Astroparticle Physics'' research project.

\appendix
\section{Estimate of $\beta\pm\sigma_\beta$}

In this Appendix we derive Eq.~(\ref{beta+-}). We make use of the
fact that, since
\begin{equation}\label{Ntot1}
  N=\int_0^T dt\,R(t)=N_S+N_B
\end{equation}
and
\begin{equation}\label{Ntot2}
  N=\int_0^T dt\,\sum_{i=1}^{N_S+N_B}\delta (t-t_i)
\end{equation}
(where $\delta$ is a Dirac delta), we can formally write
\begin{equation}\label{dN}
  dN = R(t) dt = \sum_{i=1}^{N_S+N_B}\delta (t-t_i)\, dt\ .
\end{equation}
This ``trick''  allows to express the continuous function $R(t)$
in terms of the discrete time series $\{ t_i
\}_{i=1,\dots,N_S+N_B}$, and to evaluate the variance of the small
number $R(t)dt$ through Poisson's statistics (see also
\cite{Faid}):
\begin{equation}\label{var}
\mathrm{var}(R(t)dt)=\mathrm{var}(dN)=dN=R(t)dt\ .
\end{equation}

Let us prove that $\beta=N_S\sigma_S/S$:
\begin{eqnarray}
\beta &=& \sum_{i=1}^{N_S+N_B} f(t_i) \\
&=& \int_0^T dt\sum_{i=1}^{N_S+N_B}f(t)\,\delta(t-t_i)\\
&=& \int_0^Tdt\,R(t)\,f(t)\\
&=&
\int_0^Tdt\,\left[S\left(1+\frac{\sigma_S}{S}f(t)\right)+B\right]f(t)\\
&=& \int_0^Tdt\,\sigma_S\,f^2(t) \\
&=& \sigma_S\, T\\
&=& \sigma_S \frac{N_S}{S}\ .
\end{eqnarray}

Finally, let us prove that, for $\sigma_S/S\ll 1$, it is
$\sigma_\beta\simeq \sqrt{N_S+N_B}$:
\begin{eqnarray}
\sigma^2_\beta &=& \mathrm{var}(\beta)\\
&=& \mathrm{var}\left(\int_0^T dt\, R(t)\, f(t)\right)\\
&=& \int_0^T dt\,f^2(t)\, R(t)\label{vardistr}\\
&=& (S+B)T +ST\frac{\sigma_S}{S}\left[\frac{1}{T}\int dt\,f^3(t)\right]\label{long}\\
&\simeq& (S+B)T \label{S+T}\\
&=&N_S+N_B
\end{eqnarray}
In deriving Eq.~(\ref{vardistr}), we have used both the
distributive property $\mathrm{var}(\sum_i a_i x_i)=\sum_i a^2_i
\mathrm{var}(x_i) $ and Eq.~(\ref{var}). In deriving
Eq.~(\ref{S+T}), we have dropped the second term on the
right-hand-side of Eq.~(\ref{long}), which is not only of
$O(\sigma_S/S)< 1$ but, in typical applications (including all
those considered in this work), is further suppressed by the small
value of the third moment of $f(t)$.

\section{Relation with the Kolmogorov-Smirnov unbinned test}

The Kolmogorov-Smirnov (KS) unbinned test can be used to compare
two distributions through the maximal difference between their
{\em cumulative\/} distributions \cite{Stat}. In our case, the two
distributions would be the total event rate $R(t)$ including time
variations [Eqs.~(\ref{RSB}) and (\ref{S(t)})] and the total event
rate excluding time variations, $R_0=S+B$. It is easy to convince
oneself that, in general, the maximal difference $d$ between the
corresponding cumulative distributions is of the kind:
\begin{equation}\label{dKS}
d = \eta \frac{\sigma_S}{S}\frac{N_S}{N_S+N_B}\ ,
\end{equation}
where $\eta$ is a dimensionless factor of $O(1)$, which depends on
the functional form of $f(t)$. The KS test then ``measures'' the
difference $d$ in units of $\kappa/\sqrt{N}$ (for large
$N=N_S+N_B$), where the number $\kappa$ depends on the confidence
level chosen \cite{Stat}. Therefore, the statistical significance
of the KS test is proportional to
\begin{equation}\label{KS}
  \frac{d \sqrt{N}}{\kappa}\propto
  \frac{\sigma_S}{S}\frac{N_S}{\sqrt{N_S+N_B}}\ .
\end{equation}
The comparison of the above result with Eq.~(\ref{nsdfull})
suggests that the statistical significance of time variations
should scale with $(\sigma_S/S)/(N_S/\sqrt{N_S+N_B})$,
independently of the specific (unbinned) statistical test
chosen---a fact than can be useful in prospective studies of
experimental sensitivities. Different tests may differ, however,
in statistical power. In all the examples discussed in this work,
the KS test happens to be less powerful than the one proposed
through Eq.~(\ref{nsdfull}), although we cannot exclude that it
might be more powerful in other cases.


\begin{thebibliography}{99}

\bibitem{Revi}  A.B.~McDonald, C.~Spiering, S.~Schonert, E.T.~Kearns, and T.~Kajita,
                Rev.\ Sci.\ Instrum.\  {\bf 75}, 293 (2004).

\bibitem{SKde}  SK Collaboration, Y.~Fukuda {\em et al.},
                Nucl.\ Instrum.\ Meth.\ A {\bf 501}, 418 (2003).

\bibitem{SNOd}  SNO Collaboration, J.~Boger {\em et al.},
                Nucl.\ Instrum.\ Meth.\ A {\bf 449}, 172 (2000).

\bibitem{KLde}  KamLAND Collaboration, D.M.~Markoff {\em et al.},
                J.\ Phys.\ G {\bf 29}, 1481 (2003).

\bibitem{BORE}  Borexino Collaboration, G.~Alimonti {\it et al.},
                Astropart.\ Phys.\  {\bf 16}, 205 (2002).

\bibitem{Four}  G.L.~Fogli, E.~Lisi and D.~Montanino,
                Phys.\ Rev.\ D {\bf 56}, 4374 (1997).

\bibitem{Faid}  B.~Fa\"{\i}d, G.L.~Fogli, E.~Lisi and D.~Montanino,
                Astropart.\ Phys.\  {\bf 10}, 93 (1999).

\bibitem{Lomb}  SK Collaboration, J.~Yoo {\it et al.},
                Phys.\ Rev.\ D {\bf 68}, 092002 (2003).

\bibitem{Pedr}  P.C.~de Holanda and A.Yu.~Smirnov,
                hep-ph/0309299.

\bibitem{MSmy}  SK Collaboration, M.B.~Smy {\it et al.},
                Phys.\ Rev.\ D {\bf 69}, 011104 (2004).

\bibitem{Fuku}  SK Collaboration, S.~Fukuda {\it et al.},
                Phys.\ Rev.\ Lett.\  {\bf 86}, 5651 (2001).

\bibitem{SNOC}  SNO Collaboration, Q.R.~Ahmad {\it et al.},
                Phys.\ Rev.\ Lett.\  {\bf 87}, 071301 (2001).

\bibitem{SNOD}  SNO Collaboration, Q.R.~Ahmad {\it et al.},
                Phys.\ Rev.\ Lett.\  {\bf 89}, 011301 (2002).

\bibitem{SNOA}  SNO Collaboration, Q.R.~Ahmad {\it et al.},
                Phys.\ Rev.\ Lett.\  {\bf 89}, 011302 (2002).

\bibitem{SNON}  SNO Collaboration, S.N.~Ahmed {\it et al.},
                nucl-ex/0309004.

\bibitem{Ours}  G.L.~Fogli, E.~Lisi, A.~Marrone, and A.~Palazzo,
                Phys.\ Lett.\ B {\bf 583}, 149 (2004).

\bibitem{Matt}  L.~Wolfenstein,
                Phys.\ Rev.\ D {\bf 17}, 2369 (1978);
                S.P.~Mikheev and A.Yu.\ Smirnov,
                Yad.\ Fiz.\ {\bf 42}, 1441 (1985)
                [Sov.\ J.\ Nucl.\ Phys.\ {\bf 42}, 913 (1985)].

\bibitem{Petc}  A.~Bandyopadhyay, S.~Choubey, S.~Goswami, S.T.~Petcov, and D.P.~Roy,
                Phys.\ Lett.\ B {\bf 583}, 134 (2004).

\bibitem{Grah}  K.~Graham for the SNO Collaboration, talk given at
                {\em NOON 2004}, 5th International Workshop on
                Neutrino Oscillations and their OrigiN (Tokyo,
                Japan, 2004); website:
                www-sk.icrr.u-tokyo.ac.jp/noon2004~.

\bibitem{Corr}  G.L.~Fogli, E.~Lisi, and D.~Montanino,
                Phys.\ Lett.\ B {\bf 434}, 333 (1998).

\bibitem{Kras}  J.N.~Bahcall, P.I.~Krastev, and A.Yu.~Smirnov,
                Phys.\ Rev.\ D {\bf 62}, 093004 (2000);
                {\em ibidem\/} {\bf 63}, 053012 (2001).

\bibitem{Land}  KamLAND Collaboration, K.~Eguchi {\it et al.},
                Phys.\ Rev.\ Lett.\  {\bf 90}, 021802 (2003).

\bibitem{Grat}  C.~Bemporad, G.~Gratta, and P.~Vogel,
                Rev.\ Mod.\ Phys.\  {\bf 74}, 297 (2002).

\bibitem{Hort}  G.A.~Horton-Smith for the KamLAND Collaboration,
                talk given at {\em NO-VE 2003}, 2nd International
                Workshop on Neutrino Oscillations in Venice
                (Venice, Italy, 2003); website: axpd24.pd.infn.it/NO-VE/NO-VE.html

\bibitem{Fior}  F.~Mantovani, L.~Carmignani, G.~Fiorentini, and M.~Lissia,
                Phys.\ Rev.\ D {\bf 69}, 013001 (2004).

\bibitem{Hern}  J.M.\ Herndon, Proc.\ Natl.\ Acad.\ Sci.\ U.S.A.\
                {\bf 93}(2), 646 (1996); {\em ibidem\/}
                {\bf 100}(6), 3047 (2003); see also the website
                www.nuclearplanet.com~.

\bibitem{Dono}  W.F.~McDonough, {\em Compositional Models for the Earth's
                Core}, in ``Treatise on Geochemistry,'' Vol.~II,
                edited by R.W.~Carlson (Elsevier Science, Amsterdam, 2003); website
                www.treatiseongeochemistry.com~.

\bibitem{Lisi}  E.~Lisi, talk at {\em NOON 2003}, 4th International Workshop on
                Neutrino Oscillations and their OrigiN (Kanazawa,
                Japan, 2003); website:
                www-sk.icrr.u-tokyo.ac.jp/noon2003~.

\bibitem{Stat}  W.T.~Eadie, D.~Drijard, F.E.~James, M.~Roos, and
                B.~Sadoulet, {\em Statistical Methods in Experimental
                Physics\/}
                (North-Holland, Amsterdam, 1971).

\end{thebibliography}
\end{document}